\documentclass[aps,prb,twocolumn,showpacs,floatfix,superscriptaddress,amssymb,
amsmath,preprintnumbers]{revtex4}
\usepackage{graphicx}
\usepackage{dcolumn}
\usepackage{bm}
\usepackage{amsmath}
\usepackage{color}

\newcommand{\bk}{\mathbf{k}}
\newcommand{\dn}{\downarrow}
\newcommand{\half}{{\textstyle\frac{1}{2}}}
\newcommand{\pdag}{\phantom{\dagger}}
\newcommand{\up}{\uparrow}
\newcommand{\veps}{\varepsilon}

\begin{document}
\title{Spin-polarized conductance in double quantum dots: \\
Interplay of Kondo, Zeeman and interference effects}

\author{Luis G.\ G.\ V.\ Dias da Silva}
\affiliation{Instituto de F\'{\i}sica, Universidade de S\~{a}o Paulo,
C.P.\ 66318, 05315--970 S\~{a}o Paulo, SP, Brazil}
\author{E.\ Vernek}
\affiliation{Instituto de F\'{\i}sica, Universidade Federal de Uberl\^andia,
Uberl\^andia, MG 38400--902, Brazil}
\author{K.\ Ingersent}
\affiliation{Department of Physics, University of Florida, P.O.\ Box 118440,
Gainesville, Florida, 32611--8440}
\author{N.\ Sandler}
\affiliation{Department of Physics and Astronomy, and Nanoscale
and Quantum Phenomena Institute, \\ Ohio University, Athens, Ohio
45701--2979}
\affiliation{Dahlem Center for Complex Quantum Systems and Fachbereich Physik,
Freie Universit\"at Berlin, 14195 Berlin, Germany}
\author{S.\ E.\ Ulloa}
\affiliation{Department of Physics and Astronomy, and Nanoscale
and Quantum Phenomena Institute, \\ Ohio University, Athens, Ohio
45701--2979}
\affiliation{Dahlem Center for Complex Quantum Systems and Fachbereich Physik,
Freie Universit\"at Berlin, 14195 Berlin, Germany}

\date{\today}
\begin{abstract}
We study the effect of a magnetic field in the Kondo regime of a
double-quantum-dot system consisting of a strongly correlated dot (the ``side
dot'') coupled to a second, noninteracting dot that also connects two external
leads. We show, using the numerical renormalization group, that application of
an in-plane magnetic field sets up a subtle interplay between electronic
interference, Kondo physics, and Zeeman splitting with nontrivial consequences
for spectral and transport properties. The value of the side-dot spectral
function at the Fermi level exhibits a nonuniversal field dependence that can be
understood using a form of the Friedel sum rule that appropriately accounts for
the presence of an energy- and spin-dependent hybridization function.
The applied field also accentuates the exchange-mediated interdot coupling, which
dominates the ground state at intermediate fields leading to the formation of
antiparallel magnetic moments on the dots. By tuning gate voltages
and the magnetic field, one can achieve complete spin polarization of the linear
conductance between the leads, raising the prospect of applications of the device
as a highly tunable spin filter. The system's low-energy properties are
qualitatively unchanged by the presence of weak on-site Coulomb repulsion within
the second dot.
\end{abstract}


\maketitle

\section{Introduction}

Electron correlations in quantum-dot structures result in many fascinating
effects that can be probed in detail with remarkable experimental control of
system parameters.\cite{Gordon1,Marcus,Cronenwett} Perhaps one of the most
interesting regimes occurs when electrons confined in the dot acquire
antiferromagnetic correlations with electrons in the leads, giving rise to the
well-known Kondo effect.\cite{Hewson} The simplest realization of this
phenomenon in a single quantum dot is characterized by just one low-energy
scale, set by the Kondo temperature, which controls (among other features) the
width of a many-body resonance at the Fermi energy.\cite{Gordon1,Hewson}
Recent experimental\cite{DQD1,DQD2,DQD3,DQD4,DQD5,vdWiel}
studies of the Kondo effect in multiple quantum dots have revealed a complex
competition between geometry and correlations, making evident that these
structures provide a flexible setting in which to explore much novel physics.

In this context, double-quantum-dot arrangements exhibit striking
manifestations of Kondo physics,
with conductance signatures of these effects predicted to show up in realistic
experimental setups.
A telling example is the interplay of Kondo physics and quantum interference in
``side-coupled'' or ``hanging-dot'' configurations,
\cite{Apel_EPJB04,Busser,
Cornaglia,Tanaka_PRB2005,Zitko1,Tanaka2008,ZitkoFano,Tanaka2012,Irisnei2011}
leading to a variety of interesting ``Fano-Kondo" effects. \cite{SpinPolRMP}
A rather unexpected situation arises when a small, strongly interacting ``dot
1'' is connected to external leads via a large ``dot 2'' that is tuned to have
a single-particle level in resonance with the common Fermi energy of the
leads.\cite{Dias1,DiasComment,Dias2} In this configuration the Kondo resonance,
which normally has a single peak at the Fermi energy, splits into two peaks---a
behavior that can be understood as a consequence of interference between the
many-body Kondo state in dot 1 and a single-particle-like resonance that
controls (or ``filters'') its connection to the
leads.\cite{Dias1,DiasComment,Dias2,Aligia} The magnitude of the Kondo peak splitting
is determined by the balance of several important energy scales in the problem:
the width and position of the active single-particle level in dot 2; the height
of the effective single-particle resonance set by the interdot coupling; and
the many-body Kondo temperature (determined by the preceding energy scales in
combination with the dot-1 level-position and interaction strength). This
filtering of the leads preserves a fully screened Kondo ground state with a
Kondo temperature that rises with increasing interdot coupling.

In this work, we investigate the effects of an external in-plane magnetic field
on such a double-quantum dot-system in the side-dot arrangement. The
field---which introduces another energy scale, the Zeeman energy---is known to
be detrimental to the Kondo state in single-dot systems.\cite{Andrei,Gordon1,
Hofstetter,Costi00,Logan:JPCM:9713:2001} Using numerical renormalization-group
methods,\cite{NRG,Hofstetter} we study the interplay between the different
energy scales and discuss the behavior of the Kondo resonance in the presence of
competing interactions. This interplay reveals itself in the fundamental
Fermi-liquid properties of the system, such as the variation with magnetic field
$B$ at zero temperature of the Fermi-energy ($\omega=0$) value of the side-dot
spectral function $A_1(\omega,T)$. Instead of the usual monotonic
decay\cite{Costi00,Logan:JPCM:9713:2001} of $A_1(0,0)$ with increasing $B$ we
find a markedly nonuniversal behavior, where $A_1(0,0)$ passes through a
maximum at a nonzero value of the field. This effective
\textit{field-enhancement} of the Kondo spectral function is a consequence of
the side-dot geometry. The same behavior can also be understood using an
appropriate form of the Friedel sum rule, which predicts parameter- and
field-dependent phase shifts that impart the unusual nonmonotonicity to the
variation of $A_1(0,0)$ with $B$.

In addition, we show that the competition between Zeeman splitting of the dot
levels and Kondo screening results in a dominant exchange-mediated
antiferromagnetic coupling of the dots over a range of moderate magnetic fields,
before both dots become fully polarized at higher fields. Finally, we identify
signatures of the aforementioned phenomena in the transport properties.
A key result is the generation of \textit{spin-polarized currents} through the
device, which can be tuned by adjusting gate voltages to achieve \textit{total
polarization.}

The remainder of the paper is organized as follows: In Sec.\ \ref{sec:System}
we describe the effective Anderson impurity model for the double-quantum-dot
system. Section \ref{sec:SpecFeatures} presents the low-energy spectral
properties, while Sec.\ \ref{sec:GenFSR} interprets the nonuniversal behavior
of $A_1(\omega=0,T=0)$ vs $B$ in terms of the Friedel sum rule.
The transport properties, including spin polarization, are explored in Sec.\
\ref{sec:Conductance}. Concluding remarks appear in Sec.\ \ref{sec:conclusion}.

\section{Double-quantum-dot system}
\label{sec:System}

\begin{figure}
\includegraphics[width=2.6in]{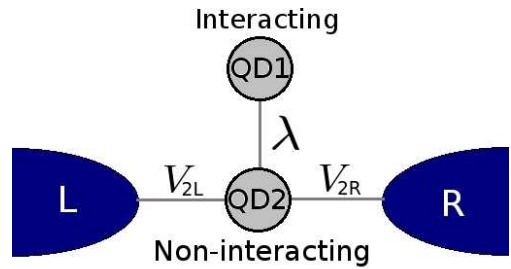}
\caption{\label{fig:model} (Color online)
Schematic representation of the side-coupled double-quantum-dot system. The dot
QD1 has a strong Coulomb interaction $U_1$ and is coupled only to the second
dot, labeled QD2. The latter dot has negligible local interactions (i.e.,
$U_2 \simeq 0$) and the energy of its active level is tuned to allow tunneling
at or near resonance with the Fermi level of the left ($L$) and right ($R$)
leads.}
\end{figure}

The system under study, which is depicted schematically in Fig.\
\ref{fig:model}, contains two quantum dots. Dot 1 has a large Coulomb repulsion
$U_1$ when its single active energy level is doubly occupied.
Dot 2 has negligible electron-electron interactions ($U_2\simeq 0$) and one
active level that can be tuned by gate voltages to be at or near resonance with
the common Fermi energy $\epsilon_F=0$ of left ($L$) and right ($R$) leads.
Electrons can tunnel between dots 1 and 2 with tunneling matrix element
$\lambda$, and between dot 2 and lead $\ell$ with tunneling matrix element
$V_{2\ell}$.

The system can be described by a variant of the two-impurity Anderson
Hamiltonian:
\begin{equation}
\label{eq:H}
 H=H_{\mathrm{dots}} + H_{\mathrm{leads}}+H_{\mathrm{hyb}},
\end{equation}
with
\begin{align}
\label{eq:H_dots}
H_{\mathrm{dots}}
& =\sum_{i=1,2} \Biggl(\sum_{\sigma} \veps_{i\sigma} n_{i\sigma}
   + U_i n_{i\up}n_{i\dn} \Biggr) \notag \\
& \qquad +\lambda\sum_{\sigma} \bigl(d_{1\sigma}^{\dag} d_{2\sigma}^{\pdag}
  + \mathrm{H.c.} \bigr), \\
\label{eq:H_leads}
H_{\mathrm{leads}}
& = \sum_{\ell=L,R} \sum_{\bk,\sigma} \veps_{\ell\bk\sigma}
  c_{\ell\bk\sigma}^{\dag} c_{\ell\bk\sigma}^{\pdag} \, ,
\end{align}
and
\begin{equation}
\label{eq:H_hyb}
H_{\mathrm{hyb}} = \sum_{\ell=L,R} V_{2\ell} \sum_{\bk,\sigma} \bigl(
 d_{2\sigma}^{\dag} c_{\ell\bk\sigma}^{\pdag}+ \mathrm{H.c.}\bigr) .
\end{equation}
Here, $d_{i\sigma}$ annihilates an electron in dot $i$ with spin $z$ component
$\half\sigma$ ($\sigma=\pm 1$ or equivalently $\up,\:\dn$) and energy
$\veps_{i\sigma}=\veps_i+\half \sigma g_i \mu_B B$ ;
$n_{i\sigma}=d_{i\sigma}^{\dag} d_{i\sigma}^{\pdag}$ is the corresponding
number operator; and $c_{\ell\bk\sigma}$ annihilates an electron in lead $\ell$
with spin $z$ component $\half\sigma$ and energy
$\veps_{\ell\bk\sigma}=\veps_{\ell\bk}+\half \sigma g_c \mu_B B$.
The magnetic field $B\hat{\mathbf{z}}$ with $B\ge 0$ is assumed to lie in the
plane of the two-dimensional electron gas in which the dots and leads are
defined, so that it produces no kinematic effects and enters only through Zeeman
level splittings. This Hamiltonian differs from a generic two-impurity Anderson
model through the absence of dot-1 hybridizations $V_{1\ell}$, a consequence of
the side-dot geometry. Throughout the greater part of the paper, we also take
$U_2=0$, a case that is particularly convenient for algebraic analysis. The
effect of nonvanishing dot-2 interactions is addressed at the end of Sec.\
\ref{sec:Conductance}.

Without loss of generality, we take all tunneling matrix elements to be real.
We consider local ($\bk$-independent) dot-lead tunneling and assume that the
dots have equal effective $g$ factors $g_1=g_2=g$, simplifications that do not
qualitatively affect the physics. The leads are taken to have featureless
band structures near the Fermi energy, modeled by the flat-top densities of
states $\rho_L(\omega)=\rho_R(\omega)=\rho(\omega)=(2D)^{-1}\Theta(D-|\omega|)$
where $D$ is the half-bandwidth and $\Theta(x)$ is the Heaviside function. The
equilibrium and linear-response properties of the system may be
calculated\cite{one-eff-band} by considering the coupling of dot 2 via
hybridization matrix element $V_2 = \sqrt{V_{2L}^2+V_{2R}^2}$ to a single
effective conduction band described by annihilation operators
$c_{\bk\sigma}= (V_{2L} c_{L\bk\sigma} + V_{2R} c_{R\bk\sigma})/V_2$ and a
density of states $\rho(\omega)$.
The Zeeman splitting of this conduction band produces only very small effects
near the band edges, so for convenience we set the bulk $g$ factor to $g_c=0$
throughout what follows.

The primary quantities of interest in this work are the retarded dot Green's
functions ${\cal G}_{i\sigma}(\omega,T)=\langle\langle d_{i\sigma}^{\pdag};
d_{i\sigma}^{\dag}\rangle\rangle_\omega$ for $i=1, \, 2$, where $\langle\langle
A;B\rangle\rangle_\omega= -i\int_0^{\infty}\langle\{A(t),B(0)\}\rangle
e^{i\omega t}dt$ and $\langle\cdots\rangle$ denotes an appropriate thermal
average.\cite{Zubarev} In particular, we are interested in the spectral
functions $A_{i\sigma}(\omega,T)=
-\pi^{-1}\mathrm{Im}\,\mathcal{G}_{i\sigma}(\omega,T)$ and the system's linear
(zero-bias) conductance, given by the Meir-Wingreen formula\cite{Meir_Wingreen}
as $G = \sum_{\sigma} G_{\sigma}$ with
\begin{equation}
\label{eq:G}
G_{\sigma}(T) = \half \, G_0 \int_{-\infty}^\infty [-\mathrm{Im} \,
  \mathcal{T}_{\sigma}(\omega,T)]\, (-\partial f/\partial\omega) \,
  d\omega,
\end{equation}
where $f(\omega,T)$ is the Fermi distribution function at temperature $T$ and
$G_0 = [2 V_{2L} V_{2R} / (V_{2L}^2 + V_{2R}^2)]^2 (2e^2 / h)$ represents the
unitary conductance of a single channel of electrons multiplied by a
factor\cite{one-eff-band,LoganPRB2009} that varies between 1 (for
$V_{2L}=V_{2R}$) and 0 (in the limit of extreme left-right asymmetry of the
dot-tunneling). In the side-connected geometry, the transmission
is\cite{Apel_EPJB04,Cornaglia,Dias2}
\begin{equation}
\label{eq:GTG22}
\mathcal{T}_{\sigma}(\omega,T)=\Delta_2 \, \mathcal{G}_{2\sigma}(\omega,T),
\end{equation}
where $\Delta_2=\pi V^2_2/2D$.
Thus,
\begin{equation}
\label{eq:G:hanging}
2G_{\sigma}(T)/G_0 = \pi \Delta_2 \int_{-\infty}^{\infty}
  A_{2\sigma}(\omega,T) \, (-\partial f/\partial\omega) \, d\omega ,
\end{equation}
which reduces at zero temperature to
\begin{equation}
\label{eq:G:hanging:T=0}
2G_{\sigma}(T=0)/G_0 = \pi \Delta_2 A_{2\sigma}(0,0).
\end{equation}

In order calculate the dot spectral functions $A_{i\sigma}(\omega,T)$ taking
full account of the electronic correlations arising from the $U_1$ term
in Eq.\ \eqref{eq:H_dots}, we employ the numerical renormalization-group (NRG)
method, performing a logarithmic discretization of the conduction band and
iteratively solving the discretized Hamiltonian. In evaluating the spectral
functions, we perform a Gaussian-logarithmic broadening of discrete poles
obtained by the procedure described in Ref.\ \onlinecite{Bulla01}. At
temperatures $T>0$ we use the density-matrix variant of the
NRG,\cite{Hofstetter} which has better spectral resolution at high frequencies
and nonzero fields.\cite{Hofstetter,NRG}
Although these schemes are not totally free from systematic errors,
\cite{BroadeningMagField} the main results of the paper do not depend crucially
on the broadening procedure.

All numerical results were obtained for a symmetric dot 1 described by
$U_1=-2\veps_1$, for dot 2 width $\Delta_2=0.02$, and for NRG discretization
parameter $\Lambda=2.5$. Except where it is stated otherwise, we consider a
strongly correlated dot 1 with $U_1=0.5$ and situations in which
a noninteracting dot 2 is tuned to be in resonance with the leads, i.e.,
$U_2=\veps_2=0$. We adopt units in which $D=\hbar=k_B=g\mu_B=1$.

\section{Spectral Properties}
\label{sec:SpecFeatures}

\begin{figure}
\includegraphics[width=3.3in]{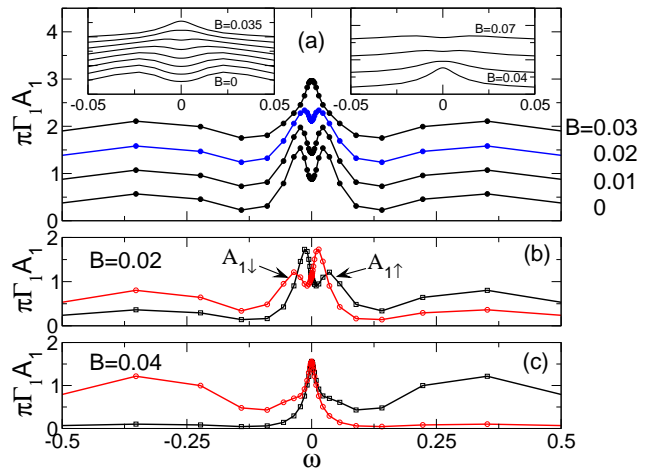}
\caption{\label{fig:A_vs_w} (Color online)
(a) Spin-averaged dot-1 spectral function $A_1$ vs frequency $\omega$ at zero
temperature for $U_1=-2\veps_1=0.5$, $\veps_2=0$, $\lambda=0.0627$, and (from
bottom to top curve, offset for clarity) $B=0$, $0.01$, $0.02$, and $0.03$.
The spectral function is multiplied by $\pi\Gamma_1$ where $\Gamma_1 =
\lambda^2/\Delta_2$. Insets: Expanded views of $A_1$ vs $\omega$ around
the Fermi level $\omega=0$ for the same system, with $B$ ranging from $0$ to
$0.035$ (bottom to top, curves offset for clarity) in steps of $0.005$ in the
left inset, and from $0.04$ to $0.07$ (bottom to top, curves offset for clarity)
in steps of $0.01$ in the right inset.
(b) Spin-up (black squares) and spin-down (red circles) dot-1 spectral functions
$A_{1\sigma}(\omega,T=0)$ for $B=0.02$ with all other parameters as in (a).
(c) Same as (b), except for $B=0.04$.}
\end{figure}

In single quantum dots, the presence of an in-plane magnetic
field \cite{Hofstetter} or connection to ferromagnetic leads\cite{Martinek}
modifies coherent spin fluctuations and weakens the Kondo effect. The
spin-averaged spectral function exhibits a Kondo-peak splitting that grows
with increasing applied field, while the value of the spectral function at the
Fermi energy decreases monotonically. In this section we investigate the
effects of a Zeeman field on the side-dot spectral function in the
double-dot system defined in Sec.\ \ref{sec:System}.

Figure \ref{fig:A_vs_w}(a) shows the spin-averaged spectral function
\begin{equation}
A_1(\omega,T) =
\half[A_{1\up}(\omega,T)+A_{1\dn}(\omega,T)]
\end{equation}
for a side-dot setup at zero temperature with $U_1=0.5$ and
$\lambda=0.0627$. The different curves, vertically offset for clarity,
correspond to four different values of $B$. For zero field (the bottom curve),
$A_{1\up}(\omega,0)=A_{1\dn}(\omega,0) =A_1(\omega,0)$, so each spin-resolved
spectral function shows a symmetric Kondo-peak splitting due to the interdot
coupling $\lambda$. With increasing $B$, the split peaks merge into a single
peak at $\omega=0$, clearly seen for $B=0.03$ (top curve). The left inset to
Fig.\ \ref{fig:A_vs_w}(a) shows in greater detail the convergence of the peaks
near the Fermi energy, with the maximum in $A_1$ vs $\omega$ at $\omega=0$
being best defined at $B=0.035$, a field where, incidentally, the absolute
value of $A_1(0,0)$ exceeds that at $B=0$ by nearly a factor of two. For
slightly larger fields, the central peak again splits into two before all
low-energy features become flattened out at fields $B\ge 0.07$ [right inset to
Fig.\ \ref{fig:A_vs_w}(a)].

The field-induced merging of the peaks in $A_1(\omega,0)$ arises from opposite
displacements of $A_{1\up}(\omega,0)$ and $A_{1\dn}(\omega,0)$ along the
$\omega$ axis. In a nonzero magnetic field, $A_{1\sigma}(\omega,T)\ne
A_{1\sigma}(-\omega,T)$ but $A_{1\up}(\omega,T)=A_{1\dn}(-\omega,T)$. This is
illustrated for $B=0.02$ in Fig.\ \ref{fig:A_vs_w}(b), which also shows that
the heights of the two peaks in each spin-resolved spectral function
$A_{1\sigma}(\omega,0)$ are no longer equal. Upon further increase in the field
to $B=0.04$ [Fig.\ \ref{fig:A_vs_w}(c)], the double-peak structure is replaced
by a single peak near $\omega=0$ in each spin-resolved spectral function. For
larger values of $B$, these peaks move away from the Fermi energy and the usual
Zeeman-splitting of the Kondo peak with decreasing amplitude becomes evident in
the spin-averaged spectral function [right inset in Fig.\ \ref{fig:A_vs_w}(a)].
This behavior can be qualitatively understood by considering the
evolution with $B$ of the level energies found\cite{Vaugier2007} in the ``atomic
limit'' $\Delta_{2} = 0$ where the dots are isolated from the leads.

\begin{figure}
\includegraphics[width=3.3in]{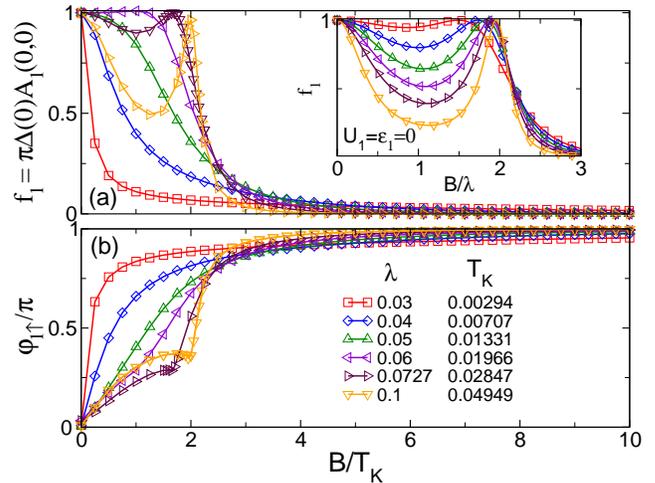}
\caption{\label{fig:A_vs_B} (Color online)
(a) Spin-averaged dot-1 spectral function at the Fermi level
$A_1(\omega\!=\!0,\,T\!=\!0)$ vs scaled magnetic field $B/T_K$ for
$U_1=-2\veps_1=0.5$, $\veps_2=0$, and six values of $\lambda$. $A_1(0,0)$ has
been multiplied by the field-dependent quantity $\pi\Delta(0)$
[Eq.\ \eqref{eq:Delta(0)}] to yield $f_1(B)$ defined in Eq.\ \eqref{eq:f_1}.
The larger $\lambda$ values produce a nonmonotonic field
variation of $A_1(0,0)$, with a peak around $B\simeq 2T_K$.
Inset: Corresponding plot for the noninteracting case $U_1=\veps_1=0$, with the
field scaled by the interdot coupling $\lambda$.
(b) Phase factor $\varphi_{1\up}=-\varphi_{1\dn}$ corresponding to the data in
(a), determined from the Friedel sum rule [Eq.\ \eqref{eq:GenFSR2}]
using the magnetization data plotted in Fig.\ \ref{fig:M_vs_B}.}
\end{figure}

We now focus on the field dependence of $A_1(\omega=0,T=0)$, a quantity that
acts as a sensitive measure of the interplay of the different energy scales in
the problem: the single-particle resonance width $\Delta_2$, the zero-field
Kondo temperature $T_K$, and the Zeeman energy $g\mu_B B$. Figure
\ref{fig:A_vs_B}(a) plots $\pi\Delta(0)\,A_1(0,0)$ vs $B/T_K$ (taking
$g\mu_B=1$) for six values of $\lambda$. The energy scale $\Delta(0)$,
introduced for normalization purposes, is defined in Eq.\ \eqref{eq:Delta(0)}
below. For now, it suffices to note that $\Delta(0)$ is proportional to
$[1+(B/2\Delta_2)^2]^{-1}$, i.e., it is a decreasing function of the field.
The figure reveals two distinct regimes of behavior: (1) For
$\lambda\lesssim 0.05$, $\pi\Delta(0)\,A_1(0,0)$ decreases monotonically from
its zero-field value $1$ over a characteristic field scale that increases with
$\lambda$ (and is not simply $T_K$, as it is in the single-dot case). (2) For
$\lambda>0.05$, $\pi\Delta(0)\,A_1(0,0)$ has a nonmonotonic variation with
increasing $B$, reaching a second maximum $\pi\Delta(0)\,A_1(0,0)=1$ at
$B=B^*\simeq 2T_K$, beyond which field it decreases.
In view of the field dependence of $\Delta(0)$, the value of
$A_1(0,0)$ at $B=B^*$ is $[1+(B^*/2\Delta_2)^2]$ times its zero-field
counterpart. The two regimes seen in Fig.\ \ref{fig:A_vs_B}(a) are in sharp
contrast with the monotonically decreasing and universal dependence of the
Fermi-energy spectral function on $B/T_K$ in the conventional single-impurity
Kondo\cite{Costi00} and Anderson\cite{Logan:JPCM:9713:2001} models. The next
section discusses these behaviors in terms of the Friedel sum rule.

\section{Friedel Sum Rule}
\label{sec:GenFSR}

In Sec.\ \ref{subsec:1imp} we review the Fermi-liquid relation known as the
Friedel sum rule\cite{Langreth1966,Hewson} that sets the Fermi-energy value of
the zero-temperature spectral function in the one-impurity Anderson model, and
write down a form of the sum rule valid for systems featuring
both a Zeeman field and nontrivial structure in the density of states. Section
\ref{subsec:2imp} shows how the variation of $A_1(\omega=0,T=0)$ in our
double-quantum-dot system can also be understood in terms of the Friedel sum
rule.

\subsection{Single Anderson impurity}
\label{subsec:1imp}

We consider a single-impurity Anderson model
\begin{align}
\label{eq:H_1imp}
H
&= \sum_{\sigma} \veps_{d\sigma} n_{d\sigma}
   + U n_{d\up}n_{d\dn}
   + \sum_{\bk,\sigma}\veps_{\bk\sigma} c_{\bk\sigma}^{\dag}
     c_{\bk\sigma}^{\pdag} \notag \\
&\qquad + \sum_{\bk,\sigma} \bigl( V_{\bk} d_{\sigma}^{\dag}
   c_{\bk\sigma}^{\pdag} + \mathrm{H.c.}\bigr) ,
\end{align}
where $\veps_{d\sigma}=\veps_d+\half\sigma g\mu_B B$ and
$\veps_{\bk\sigma}=\veps_{\bk} + \half\sigma g_c\mu_B B$.
The conduction-band dispersion $\veps_{\bk}$ and the hybridization $V_{\bk}$
enter the impurity properties only in a single combination: the zero-field
hybridization function $\Delta_0(\omega)=\pi\sum_{\bk} |V_{\bk}|^2
\delta(\omega-\veps_{\bk})$.
We denote the fully interacting retarded impurity Green's for this problem by
\begin{equation}
\label{G_d,sigma}
\mathcal{G}_{d\sigma}(\omega,T) =
\langle\langle d_{\sigma}^{\pdag};d_{\sigma}^{\dag}\rangle\rangle_{\omega}
= \frac{1}{\omega + i0^+ -\veps_{d\sigma}-\Sigma_{d\sigma}(\omega,T)} \, ,
\end{equation}
where $\Sigma_{d\sigma}(\omega,T)$ is the retarded impurity self-energy.

In the conventional Anderson model, where the hybridization function is
assumed to take a flat-top form $\Delta_0(\omega)=\Gamma \Theta(D-|\omega|)$,
the Friedel sum rule relates the Fermi-energy value of the zero-temperature,
zero-field impurity spectral function $A_d(\omega,0) \equiv
A_{d\sigma}(\omega,0)= -\pi^{-1}\mathrm{Im}\,\mathcal{G}_{d\sigma}(\omega,0)$
to the average impurity occupancy
$\langle n_d\rangle=\langle n_{d\up}\rangle+\langle n_{d\dn}\rangle$ as
\begin{equation}
\label{eq:FSRconstHyb:B=0}
\pi \Gamma A_d(0,0) = \sin^2\left(\frac{\pi}{2}\langle n_d\rangle\right) \, .
\end{equation}
In the wide-band limit where $D$ greatly exceeds all other energy scales in
the problem, Eq.\ \eqref{eq:FSRconstHyb:B=0} has been extended \cite{WrightPRB2011}
to show that in a Zeeman field $B$, the spin-averaged impurity spectral function
$A_d(\omega,T)=\half[A_{d\up}(\omega,T)+A_{d\dn}(\omega,T)]$ satisfies
\begin{equation}
\label{eq:GenFSR:flat}
\pi \Gamma A_d(0,0) = \half \bigl[ 1 - \cos(\pi\langle n_d\rangle) \,
  \cos\bigl(2\pi M_d)\bigr],
\end{equation}
where $M_d(B)=\half (\langle n_{d\up}\rangle-\langle n_{d\dn}\rangle)$
is the impurity magnetization in units of $g\mu_B$.

Our goal is to extend Eqs.\ \eqref{eq:FSRconstHyb:B=0} and
\eqref{eq:GenFSR:flat} to allow for finite values of $D$ and any form of
$\Delta_0(\omega)$. One can show\cite{Hewson,DiasComment,Vaugier2007,Aligia,
LoganPRB2009}
that provided the system is in a Fermi-liquid regime [where the imaginary part
of $\Sigma_{d\sigma}(\omega,T=0)$ varies as $\omega^2$ for $\omega\to 0$], the
spin-resolved spectral functions at zero temperature satisfy
\begin{equation}
\label{eq:FSRNonconstHyb}
\pi \Delta_{\sigma}(0)\,A_{d\sigma}(0,0)
  = \sin^2\bigl(\pi\langle n_{d\sigma}\rangle + \varphi_{\sigma}\bigr),
\end{equation}
where $\Delta_{\sigma}(\omega)=\Delta_0(\omega-\half\sigma g_c\mu_B B)$ and
\begin{equation}
\label{eq:FSRphase}
\varphi_{\sigma} = \mathrm{Im} \int_{-\infty}^{0}
  \frac{\partial \Sigma_{d\sigma}^0(\omega,T=0)}{\partial \omega} \,
  \mathcal{G}_{d\sigma}(\omega,T=0) \; d\omega
\end{equation}
is a spin-dependent phase shift. In Eq.\ \eqref{eq:FSRphase},
$\mathcal{G}_{d\sigma}(\omega,T)$ is the fully interacting retarded impurity
Green's function specified in Eq.\ \eqref{G_d,sigma}, but
$\Sigma_{d\sigma}^0(\omega,T)$ is the retarded impurity self-energy for the
noninteracting system [Eq.\ \eqref{eq:H_1imp} with $U=0$], which satisfies
$\mathrm{Im}\,\Sigma_{d\sigma}^0(\omega,T)=-\Delta_{\sigma}(\omega)$.

In situations where $\Delta_{\up}(\omega)\ne\Delta_{\dn}(\omega)$, it is
convenient to focus on a dimensionless, hybridization-weighted average of
the spin-resolved spectral functions:
\begin{equation}
\label{eq:F}
F(\omega,T) = \frac{\pi}{2} \sum_{\sigma} \Delta_{\sigma}(\omega)
  A_{d\sigma}(\omega,T) .
\end{equation}
In terms of this quantity, the linear conductance is
\begin{equation}
\label{G_1imp}
G(T) = G_0 \int_{-\infty}^{\infty} F(\omega,T) \,
  (-\partial f/\partial\omega) \, d\omega
\end{equation}
with a zero-temperature limit
\begin{equation}
\label{G_1imp:T=0}
G(T=0) = G_0 \, F(0,0).
\end{equation}
Here, $G_0 = [2 V_L V_R / (V_L^2 + V_R^2)]^2 (2e^2 / h)$ is the maximum
possible conductance through the dot for hybridizations $V_L$ and $V_R$ with
the left and right leads, respectively. We note that the hybridization-weighted,
spin-averaged spectral function reduces to
$F(\omega,T)=\pi\Delta(\omega)\,A_d(\omega,T)$  for (i) all values of $\omega$
in zero magnetic field, and (ii) at $\omega=0$ for any field $B$ such that
$\Delta_0\bigl(\half g_c\mu_B B\bigr)=\Delta_0\bigl(-\half g_c\mu_B B\bigr)$.

Inserting Eq.\ (\ref{eq:FSRNonconstHyb}) into Eq.\ \eqref{eq:F}, rewriting
$\langle n_{d\sigma}\rangle=\half\langle n_d\rangle+\sigma M_d$, and defining
$\varphi_{\pm} = \varphi_{\up} \pm \varphi_{\dn}$, one obtains
\begin{equation}
\label{eq:GenFSR}
F(0,0) = \half \bigl[ 1 - \cos(\pi\langle n_d\rangle+\varphi_+) \,
  \cos\bigl(2\pi M_d+\varphi_-) \bigr] .
\end{equation}
This form of the Friedel sum rule relates the value of the
hybridization-weighted spin-averaged spectral function at $\omega=0$ and $T=0$
to the impurity occupancy, the impurity magnetization, and spin-dependent phase
factors that account for the energy dependence of the hybridization function.
The right-hand side of Eq.\ \eqref{eq:GenFSR} has a maximum possible value of
1, implying through Eq.\ \eqref{G_1imp:T=0} that $G(T=0)\le G_0$, as one would
expect for a problem with a single transmission mode in the left and right
leads.

In general, each of the phase factors $\varphi_{\up}$ and $\varphi_{\dn}$ has
a complicated dependence on
$\Delta(\omega)$, the
impurity parameters $U$ and $\veps_d$, and the magnetic field $B$. This makes
it highly improbable that for a generic choice of model parameters there exists
a value of $B$ for which the system satisfies the requirements
\begin{equation}
\label{eq:unitary_conditions}
\cos(\pi\langle n_d\rangle+\varphi_+)=
-\cos\bigl(2\pi M_d+\varphi_-)=\pm 1
\end{equation}
for achieving $F(0,0)=1$ and, hence, a maximum conductance $G(T=0)=G_0$.

However, under conditions where both the impurity and the conduction
band exhibit particle-hole symmetry, the Hamiltonian \eqref{eq:H} is invariant
under the transformation $d_{\sigma}^{\pdag}\to -d_{-\sigma}^{\dag}$,
$c_{\bk\sigma}^{\pdag}\to c_{\bk,-\sigma}^{\dag}$, $\veps_{\bk}\to-\veps_{\bk}$.
This invariance leads to the relations $\Delta_{\up}(\omega)=\Delta_{\dn}(-\omega)$,
$\Sigma_{d\up}^0(\omega,T) = -[\Sigma_{d\dn}^0(-\omega,T)]^*$ and
$\mathcal{G}_{d\up}(\omega,T)=-[\mathcal{G}_{d\dn}(-\omega,T)]^*$, which
in turn imply that $A_{d\up}(\omega,T) = A_{d\dn}(-\omega,T)$ and
$\varphi_{\up} =-\varphi_{\dn}$ (or $\varphi_+ = 0$). Since particle-hole
symmetry also ensures $\Delta_0(\omega)=\Delta_0(-\omega)$ and
$\langle n_d\rangle=1$, it follows that $\Delta_{\up}(0)=\Delta_{\dn}(0)$ and
 the Friedel sum rule reduces to
\begin{equation}
\label{eq:GenFSR:symm}
\pi \Delta(0) \, A_d(0,0) = \cos^2\bigl(\pi M_d+\varphi_{\up}).
\end{equation}
In situations described by Eq.\ \eqref{eq:GenFSR:symm}, the conductance will
reach its maximum possible value $G_0$ whenever $(\pi M_d+\varphi_{\up})/\pi$
equals an integer. It is much more likely that this single
condition can be met at some value of $B$ than that a system away from
particle-hole symmetry can be tuned to satisfy both parts of
Eq.\ \eqref{eq:unitary_conditions}.

The conventional flat-top hybridization function $\Delta_0(\omega)=
\Gamma\Theta(D-|\omega|)$ is not only particle-hole symmetric, but
yields vanishingly small values of $\varphi_{\sigma}$, thereby
simplifying Eq.\ \eqref{eq:GenFSR} to the previously derived\cite{WrightPRB2011}
Eq.\ \eqref{eq:GenFSR:flat}. One expects $|M_d(B)|$ to be an increasing function
of $B$ with a limiting value $|M_d(B\to\infty)|=\half$, and therefore [via Eq.\
\eqref{eq:GenFSR:flat}] both $A_d(0,0)$ and
$G(T=0)$ should decrease monotonically with increasing $B$.

\subsection{Double quantum dots}
\label{subsec:2imp}

We now return to the double-quantum-dot setup defined in Eq.\ \eqref{eq:H}.
It has been shown\cite{Dias1} that for the special case $U_2=0$, the properties
of dot 1 are identical to those of the impurity in a single-impurity Anderson
model [Eq.\ \eqref{eq:H_1imp}] with $U=U_1$, $\veps_d=\veps_1$, and a zero-field
hybridization function
\begin{equation}
\label{eq:Delta_0}
\Delta_0(\omega) = \pi\lambda^2 \rho_2(\omega),
\end{equation}
where
\begin{equation}
\label{eq:rho_2}
\rho_2(\omega) = \frac{1}{\pi}
  \frac{\Delta_2}{(\omega-\veps_2)^2+\Delta_2^2}
\end{equation}
describes a unit-normalized Lorentzian resonance of width $\Delta_2$ [defined
after Eq.\ \eqref{eq:GTG22}] centered on energy $\omega=\veps_2$.

In a Zeeman field $B$, where the spin-dependent hybridization
function of the effective one-impurity problem is
\begin{equation}
\label{eq:Delta_sigma}
\Delta_{\sigma}(\omega) = \Delta_0\bigl(\omega-\half\sigma g\mu_B B\bigr),
\end{equation}
a quantity of interest is
\begin{equation}
\label{eq:f_1}
f_1(B) = \frac{\pi}{2} \sum_{\sigma} \Delta_{\sigma}(0) \, A_{1\sigma}(0,0),
\end{equation}
the value of the hybridization-weighted spin-averaged dot-1 spectral function
at $\omega=T=0$. For the resonant case $\veps_2=0$ considered in Figs.\
\ref{fig:A_vs_w} and \ref{fig:A_vs_B},
\begin{equation}
\label{eq:Delta(0)}
\Delta_{\up}(0)=\Delta_{\dn}(0) \equiv
  \Delta(0) = \frac{\Delta_0(0)}{1 + (B/2\Delta_2)^2}
\end{equation}
in units where $g\mu_B = 1$. Taking into account also the particle-hole
symmetry present for $\veps_1=-\half U_1$ and $\veps_2=0$, the Friedel sum
rule [Eq.\ \eqref{eq:GenFSR:symm}] gives (after translation
back into the variables of the double-dot problem)
\begin{equation}
\label{eq:GenFSR2}
f_1(B) = \cos^2 (\pi M_1+\varphi_{1\up}),
\end{equation}
where $M_i=\frac{1}{2}(\langle n_{i\up}\rangle-\langle n_{i\dn}\rangle)$ is
the magnetic moment on dot $i$, and
\begin{equation}
\label{eq:FSRphase2}
\varphi_{1\sigma} = \mathrm{Im} \int_{-\infty}^{0}
  \frac{\partial \Sigma_{1\sigma}^0(\omega,T=0)}{\partial \omega} \,
  \mathcal{G}_{1\sigma}(\omega,T=0) \; d\omega
\end{equation}
with $\mathrm{Im}\,\Sigma_{1\sigma}^0(\omega,T)=-\Delta_{\sigma}(\omega)$.

\begin{figure}
\includegraphics[width=3.3in]{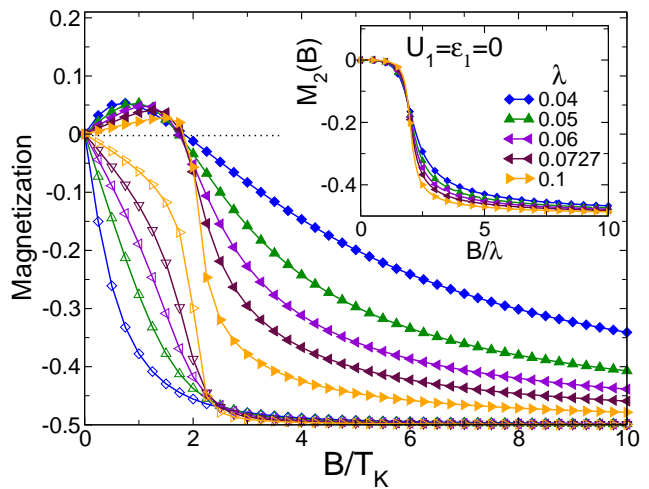}
\caption{\label{fig:M_vs_B} (Color online)
Magnetization of dot 1 (empty symbols) and dot 2 (filled symbols) vs scaled
magnetic field $B/T_K$ at zero temperature for the same parameters as in the
main panels of Fig.\ \ref{fig:A_vs_B}. The dot-1 magnetization $M_1$ decreases
monotonically from zero over a characteristic field scale that grows with
$\lambda$ and approaches $2T_K$ for sufficiently large interdot couplings. The
dot-2 magnetization $M_2$ is of opposite sign to $M_1$ for $B\lesssim 2T_K$,
pointing to the dominance of the antiferromagnetic interdot exchange
interaction over this field range. Both dots become fully polarized
antiparallel to the field for $B \gg 2T_K$. Inset: $M_2$ vs $B/\lambda$ for the
noninteracting system with the same parameters as in the inset of Fig.\
\ref{fig:A_vs_B}(a). In contrast to the interacting case, $M_2$ decreases
monotonically from zero with increasing field.}
\end{figure}

Figure \ref{fig:M_vs_B} shows the variation of $M_1$ with $B$ for the same
model parameters used in Fig.\ \ref{fig:A_vs_w}. As expected, $M_1$
decreases monotonically from zero over a field scale that grows with $\lambda$.
For small $\lambda$, this scale is identical to that characterizing the initial
decrease of $f_1$ from 1 [see Fig.\ \ref{fig:A_vs_B}(a)], while for larger
$\lambda$, $|M_1|$ grows on the scale $B^*$ of the second peak in $f_1(B)$. In
all cases, dot 1 is essentially fully polarized for $B\gtrsim 2T_K$. That the
monotonic evolution of $M_1$ does not accompany a  monotonic decrease in
$f_1(B)$ is an indication of the importance of the phase factor
$\varphi_{1\up}$ on the right-hand side of Eq.\ \eqref{eq:GenFSR2}.

It is difficult to evaluate $\varphi_{1\sigma}$ directly from Eq.\
\eqref{eq:FSRphase} using the NRG because this task requires accurate
determination of both the real and imaginary parts of
$\mathcal{G}_{\sigma}(\omega,0)$ for all $\omega<0$, whereas the NRG is
well-suited only to compute $\mathrm{Im}\,\mathcal{G}_{\sigma}(\omega,0)$
for $|\omega|\ll D$. At particle-hole symmetry, however, one can use Eq.\
\eqref{eq:GenFSR2} to work backward from the NRG values of $f_1$ and $M_1$ to
find $\varphi_{1\up}=-\varphi_{1\dn}$. Fig.\ \ref{fig:A_vs_B}(b) plots the phase
obtained in this manner from the data in Figs.\ \ref{fig:A_vs_B}(a) and
\ref{fig:M_vs_B}. For all values of $\lambda$, $\varphi_{1\up}$ is zero at
$B=0$ (as expected) and approaches $\pi$ at large field values. For larger
values of $\lambda$, $\varphi_{1\up}$ shows a pronounced kink at $B=B^*$. This
kink is related, via Eq.\ \eqref{eq:GenFSR2}, to the peak in $f_1(B)$ at $B^*$,
since $M_1(B)$ is a smooth function of $B$ (as shown in Fig.\ \ref{fig:M_vs_B}).

Figure \ref{fig:M_vs_B} also plots the field dependence of the dot-2
magnetization. The fact that $M_2$ is of \textit{opposite} sign to $M_1$ for
$B\lesssim 2T_K$ indicates that the interactions in dot 1 combine with the
interdot hopping to yield a dominant antiferromagnetic interdot exchange
interaction. Over this range of $B$, it appears that the system minimizes its
energy by first aligning the partially Kondo-screened magnetic moment of the
strongly interacting dot 1 along the direction favored by the field, and then
orienting the less-developed moment on dot 2 to minimize the interdot exchange
energy even at a cost in Zeeman energy. The data show that this tendency
becomes weaker for stronger interdot couplings, presumably because the interdot
exchange $\sim \lambda^2$ grows more slowly than the energy scale $T_K$ for
breaking the Kondo singlet. For all values of $\lambda$, once $B\gtrsim 2T_K$,
the Zeeman field has largely destroyed the Kondo effect, and both dots are
fully polarized for $B\gg 2T_K$.

One can gain further insight into the results presented in Figs.\
\ref{fig:A_vs_B} and \ref{fig:M_vs_B} by considering the limit where both
dots are noninteracting. Equations \eqref{eq:GenFSR} and \eqref{eq:GenFSR2}
hold equally well for interacting and noninteracting problems. However, the
case $U_1=U_2=0$ offers the advantage that $A_1(0,0)$ can also be calculated
directly from the imaginary part of
\begin{equation}
\label{eq:G^0}
\mathcal{G}_{1\sigma}^0(\omega,T) = \frac{1}{\omega + i0^+ - \veps_{1\sigma}
  -\Sigma_{\sigma}^0(\omega,T)},
\end{equation}
where at zero temperature the noninteracting self-energy is
\begin{equation}
\label{eq:Sigma^0}
\Sigma_{\sigma}^0(\omega,0) =
  [(\omega-\veps_{2\sigma})/\Delta_2 - i]\Delta_{\sigma}(\omega),
\end{equation}
giving
\begin{equation}
A_{1\sigma}(0,0) = \frac{1}{\pi} \, \frac{\Delta_{\sigma}(0)}
  {[\veps_{2\sigma}\Delta_{\sigma}(0)/\Delta_2 - \veps_{1\sigma}]^2
  +[\Delta_{\sigma}(0)]^2}.
\end{equation}
The hybridization-weighted spin-average of $A_{1\sigma}(0,0)$ satisfies
\begin{equation}
\label{eq:A:nonint}
f_1(B) = \frac{1}{2} \sum_{\sigma} \frac{1}{1 + (e_{2\sigma} - e_{1\sigma})^2},
\end{equation}
where $e_{1\sigma}=\veps_{1\sigma}/\Delta_{\sigma}(0)$ and
$e_{2\sigma}=\veps_{2\sigma}/\Delta_2$. It should be noted that
$e_{1\sigma}$ depends on $B$ both through the Zeeman shift of $\veps_1$
and the value of $\Delta_{\sigma}(0)=\Delta_0(-\half\sigma B)$.
From Eq.\ \eqref{eq:A:nonint} it is apparent that $f_1(B)$ attains its maximum
value of $1$ only if $e_{2\sigma}=e_{1\sigma}$ for both spin orientations, a
condition that can be satisfied only for $\veps_1=\veps_2=0$ and either $B=0$
or (if $\lambda>\Delta_2$) $B=B^*=2\sqrt{\lambda^2-\Delta_2^2}$. For
$\veps_1\ne 0$ and/or $\veps_2\ne 0$, $f_1(B)$ may have zero, one or two maxima
at nonzero fields, but $f_1<1$ for all $B$. These observations are
consistent with the conclusion drawn from the Friedel sum rule
that $f_1=1$ is likely to be achieved only under conditions of strict
particle-hole symmetry.

The inset of Fig.\ \ref{fig:A_vs_B}(a) illustrates the field variation of
$f_1$ for the particle-hole-symmetric case $U_1=\veps_1=\veps_2=0$, with
all other parameters as in the main panel. For each of the $\lambda$ values
illustrated (all of which lie in the range $\lambda>\Delta_2$), $f_1$
reaches 1 at a magnetic field consistent with the value $B^*$ derived in the
previous paragraph. Note that $B^*$ approaches $2\lambda$ from below in the
limit of strong interdot coupling. The inset of Fig.\ \ref{fig:M_vs_B} plots
$M_2$ vs $B$ for the same noninteracting cases. For each $\lambda$ value,
$|M_2|$ shows a purely monotonic field variation, with a rather sudden increase
around $B\simeq 2\lambda$, a behavior that is mimicked in the interacting
system for $B \simeq 2T_K$, especially at large interdot coupling $\lambda$.
The variation of the interacting $f_1$ and $M_2$ for $B\gtrsim 2T_K$ seen in
the main panels of Figs.\ \ref{fig:A_vs_B}(a) and \ref{fig:M_vs_B}, particularly
for the larger values of $\lambda$, may perhaps be interpreted as a many-body
analog of the noninteracting behavior in the insets, with $T_K$ serving as a
renormalized value of the single-particle scale $\lambda$.

\section{Electrical Conductance}
\label{sec:Conductance}

While the spectral functions discussed in the preceding sections are difficult
to access directly in experiments, they may be probed indirectly through
transport measurements. In this section, we show
that the zero-bias electrical conductance through the double-dot device contains clear
signatures of the nonuniversal variation of $\pi\Delta(0)\,A_1(0,0)$ with
applied field. In particular, we demonstrate the feasibility of generating
currents through the system that are strongly or even completely spin-polarized.

Although the linear conductance is given most compactly by Eq.\
\eqref{eq:G:hanging}, it is also useful to express $G$ in terms of the Green's
function for the interacting dot 1 by combining Eq.\ \eqref{eq:G} with a
generalization of Eq.\ (6) in Ref.\ \onlinecite{Dias2} to include the Zeeman
field:
\begin{align}
\label{eq:ImT}
\!\!\!\!-&\mathrm{Im}\,\mathcal{T}_{\sigma}(\omega,T) \notag \\
& \;\; = [1-2\pi\Delta_2\rho_{2\sigma}(\omega)] \,
  \pi \Delta_{\sigma}(\omega) \, A_{1\sigma}(\omega,T)
  + \pi\Delta_2\rho_{2\sigma}(\omega) \notag \\
& \qquad + 2\pi(\omega-\veps_{2\sigma})\, \rho_{2\sigma}(\omega) \,
  \Delta_{\sigma}(\omega) \: \mathrm{Re}\, \mathcal{G}_{1\sigma}(\omega,T) ,
\end{align}
where $\rho_{2\sigma}(\omega) = \rho_2(\omega-\half\sigma g\mu_B B)$, with
$\rho_2(\omega)$ and $\Delta_{\sigma}(\omega)$ as defined in Eqs.\
\eqref{eq:rho_2} and \eqref{eq:Delta_sigma}, respectively. The term
$\pi\Delta_2\rho_{2\sigma}(\omega)$ describes the bare transmission through
dot 2 in the absence of dot 1, while the remaining terms represent additional
contributions arising from conductance paths that include dot 1. In the special
case $\lambda=0$ where the latter contributions necessarily vanish, the
zero-temperature conductance reduces to
\begin{equation}
\label{eq:G:dot-2-only}
G_{\text{one-dot}}(T=0)
= \frac{G_0}{2} \sum_{\sigma} \frac{1}{1+e_{2\sigma}^2} ,
\end{equation}
where $e_{2\sigma}$ is defined after Eq.\ \eqref{eq:A:nonint}.

\subsection{Zero temperature}

\begin{figure}
\includegraphics[width=3.3in]{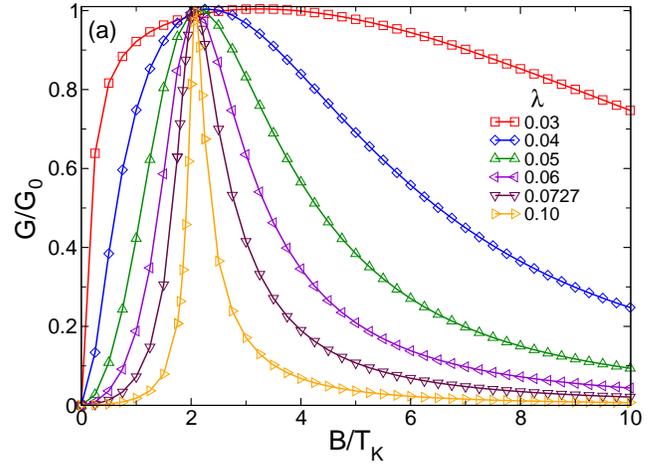}
\caption{\label{fig:G_vs_B} (Color online)
Linear conductance $G$ vs scaled magnetic field $B/T_K$ at zero temperature for
the same parameters as in the main panels of Fig.\ \ref{fig:A_vs_B}. $G$ rises
from zero over the same characteristic field scale as governs the rise of
$|M_1|$ in the main panel of Fig.\ \ref{fig:M_vs_B}.}
\end{figure}

Figure \ref{fig:G_vs_B} plots the zero-temperature linear conductance $G$ as a
function of scaled field $B/T_K$ for the same parameters used in Fig.\
\ref{fig:A_vs_B}. For the case $\veps_2=0$ considered here, the conductance of
dot 2 alone, $G_{\text{one-dot}}(T=0)=G_0[1+(B/2\Delta_2)^2]^{-1}$,
decreases monotonically from $G_0$ as the Zeeman field detunes the
dot level from the Fermi energy of the leads. For any $\lambda \neq 0$ and
$B=0$, Kondo correlations in dot 1 produce zero conductance through the
double-dot system.\cite{Dias2} Figure \ref{fig:G_vs_B} shows that
with increasing field, the double-dot conductance initially increases, then
peaks at its maximum possible value $G=G_0$ for a field value $B^{**}$ that
for large $\lambda$ approaches $2 T_K$ from above, and finally drops back
toward zero for $B\gg B^{**}$. The field $B^{**}$ is distinct from that
characterizing the peak in $\pi\Delta(0)\,A_1(0,0)$. In general
$B^*<2T_K<B^{**}$, but these three scales converge for $\lambda\gg \Delta_2$.

The initial rise in $G$ with increasing field can be attributed to the
progressive suppression of the Kondo effect allowing dot 1 to become partially
polarized and reducing the destructive interference between the Kondo resonance
and the dot-2 resonant state. This change takes place---in agreement with the
evolution seen in $\pi\Delta(0)\,A_1(0,0)$ and $M_1$---over a field scale that
increases with $\lambda$ but is not just a constant multiple of $T_K$. By the
point that the conductance reaches its peak at $B=B^{**}$, the interchannel
interference is clearly constructive since Eq.\ \eqref{eq:G:dot-2-only} would
predict a much lower conductance for dot 2 alone. At still larger fields, the
destruction of the Kondo resonance becomes complete and the dot-2 resonance is
shifted far from the Fermi level, leading to a decrease of the conductance.

\begin{figure}[t]
\includegraphics[width=3.3in]{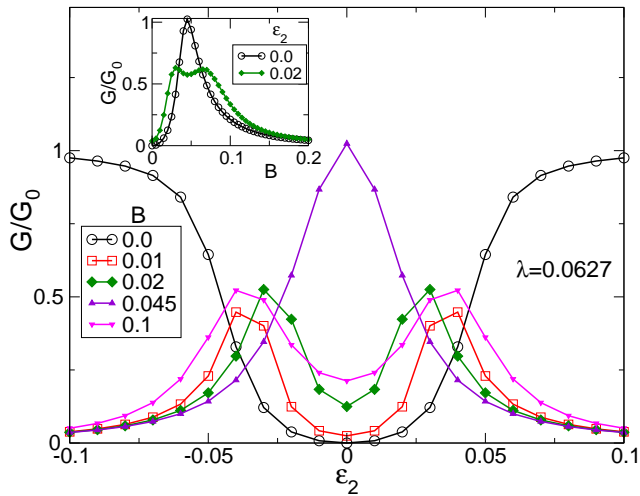}
\caption{\label{fig:CondTeq0} (color online)
Linear conductance $G$ vs dot-2 level energy $\veps_2$ at zero temperature
for $U_1=-2\veps_1=0.5$, $\lambda=0.0627$, and five different magnetic field
values. The conductance is symmetric about the point $\veps_2=0$ of
particle-hole symmetry. In nonzero fields, $G$ peaks at some $|\veps_2|\ne 0$,
apart from the special case $B=B^{**}\simeq 0.045\simeq 2T_K$, for which the
conductance is maximal at $\veps_2=0$ (as already seen in Fig.\
\ref{fig:G_vs_B}). Inset: Conductance vs magnetic field $B$ for $\veps_2=0$
and $0.02$.}
\end{figure}

Figure \ref{fig:CondTeq0} illustrates aspects of the transport away from
particle-hole symmetry. The main panel shows the variation of the $T=0$ linear
conductance at several different fixed magnetic fields as the value of
$\veps_2$ is swept by varying the voltage on a plunger gate near dot 2. For
$B=0$, the conductance increases from zero at $\veps_2=0$ and approaches $G_0$
for $|\veps_2|\gg\Delta_2$ as the dot-2 resonance is tuned away from the Fermi
energy, thereby permitting perfect conduction through the Kondo many-body
resonance. For fixed $B>0$, competition between Zeeman splitting of the dot-2
resonance and partial destruction of the Kondo effect leads in most cases to an
initial rise in $G$ for small $|\veps_2|$ followed by a fall-off at larger
$|\veps_2|$. As the magnetic field increases from zero, the conductance peaks
initially move to smaller $|\veps_2|$, then merge into a single peak at $G=G_0$
for $B=B^{**}\simeq 0.045$ for the case $\lambda=0.0627$ here, before
separating and moving to larger $|\veps_2|$ as $B$ moves to still higher
values. Thus $B^{**}$ can in principle be located as the only field at which
$G$ has a single peak vs $\veps_2$.

The inset to Figure \ref{fig:CondTeq0} compares the field variation of
$G(T=0)$ for $\veps_2=0$ and for $\veps_2 = 0.02$. It is only in the
former case (i.e., under conditions of strict particle-hole symmetry), that
the conductance has a single peak vs $B$ and attains $G=G_0$, whereas
for $\veps_2\ne 0$ one finds a pair of peaks at $G<G_0$. The presence of
a single peak under field sweeps can therefore be used to identify
the particle-hole-symmetric point in experiments.

To better understand these results, we again turn to the noninteracting case
$U_1=0$, where the linear conductance can be calculated by substituting
the noninteracting Green's function given by Eqs.\ \eqref{eq:G^0} and
\eqref{eq:Sigma^0} for the full Green's function $\mathcal{G}_{\sigma}$ in
Eq.\ \eqref{eq:ImT}. At $T=0$, this results in a conductance
contribution
\begin{equation}
\label{eq:G:T=0:nonint}
G = \sum_{\sigma} G_{\sigma} = \half\, G_0 \sum_{\sigma} \frac{e_{1\sigma}^2}
  {[1+e_{2\sigma}^2] [1+(e_{2\sigma}-e_{1\sigma})^2]} \,
\end{equation}
where $e_{1\sigma}$ and $e_{2\sigma}$ are defined after Eq.\
\eqref{eq:A:nonint}. Equation \eqref{eq:G:T=0:nonint} correctly reduces
to Eq.\ \eqref{eq:G:dot-2-only} in the limit $|\veps_1|\to\infty$ where dot 1
can play no role in the conductance.
At particle-hole symmetry ($\veps_1=\veps_2=0$), Eq.\ \eqref{eq:G:T=0:nonint}
gives
\begin{equation}
G = G_0 \frac{[B/2\Delta(0)]^2}
  {[1+(B/2\Delta_2)^2]\{1+[B/2\Delta_2 - B/2\Delta(0)]^2\}} \,
\end{equation}
which peaks at $G=G_0$ for $B = B^{**} = 2\lambda$, a characteristic
field greater than the one $B^* = 2\sqrt{\lambda^2-\Delta_2^2}$ at which
$\pi\Delta(0)\,A_1(0,0)$ reaches 1.
Since we have seen above that $G(T=0)$ for the interacting case at particle-hole
symmetry reaches $G_0$ for some $B^{**}>2T_K > B^*$, with $B^{**}\to 2T_K$
for large $\lambda$, the field dependence of the conductance reinforces the
parallels between the large-$\lambda$ interacting problem and the
noninteracting limit, with the many-body scale $T_K$ playing the role of a
renormalized $\lambda$.

An interesting feature of Eq.\ \eqref{eq:G:T=0:nonint} is that it predicts
conduction contributions $G_{\up}\ne G_{\dn}$ when particle-hole symmetry and
time-reversal symmetry are both broken. In particular, for $\veps_1>0$ (or
$\veps_1<0$), the conductance polarization measured by
\begin{equation}
\label{eq:eta}
\eta = \frac{G_{\up}-G_{\dn}}{G_{\up}+G_{\dn}}
\end{equation}
grows from $\eta=0$ for $B=0$ to reach $\eta=1$ (or $\eta=-1$) for
$B=2|\veps_1|$, at which field $\veps_{1\dn}=0$ (or $\veps_{1\up}=0$), before
decreasing toward zero for still larger fields.
By contrast, keeping $\veps_1=0$ but allowing $\veps_2\ne 0$ results in
variation of $\eta$ with field, but does not allow one to achieve perfect
polarization of the conductance.

\begin{figure}
\includegraphics[width=3.3in]{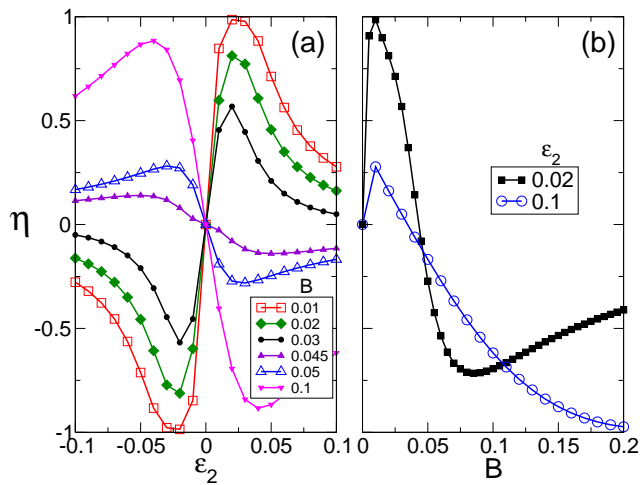}
\caption{\label{fig:PolConducTeq0} (Color online)
Conductance spin-polarization $\eta$ (a) vs dot-2 level energy $\veps_2$ at six
fixed magnetic fields $B$, and (b) vs $B$ for two values of $\veps_2$.
All data are for $U_1=-2\veps_1=0.5$, $\lambda=0.0627$, and zero temperature.
In (a), $\eta$ is odd about the point $\veps_2=0$ of particle-hole symmetry.
Complete spin polarization of the conductance is achieved in the case $B=0.01$.
Panel (b) shows a strong, nonuniversal variation of $\eta$ with $B$ for
different values of $\veps_2$.}
\end{figure}

Spin-dependent conductance is also exhibited when dot 1 has strong interactions.
Figure \ref{fig:PolConducTeq0}(a) shows the variation of $\eta$
with the dot 2 level energy $\veps_2$ in different fields $B\ne 0$ for a
symmetric dot 1 ($U_1=-2\veps_1$) and fixed $\lambda$. The conductance
spin-polarization is odd about the point $\veps_2=0$ of particle-hole symmetry
where the condition $A_{1\up}(\omega,T)=A_{1\dn}(-\omega,T)$ ensures [via
Eq.\ \eqref{eq:G:hanging}] that $\eta=0$.
For fields $B\lesssim 2T_K \simeq 0.042$, $\eta$ has the same sign as
$\veps_2$, whereas for $B\gtrsim 2T_K$, $\eta$ and $\veps_2$ have opposite
signs. For each field value, $|\eta|$ peaks at a nonzero value of $|\veps_2|$.
One sees that a field $B=0.01$ combines with a level energy $|\veps_2|\simeq
0.025$ to achieve complete destructive interference of the conduction for one
spin species, allowing passage only of a fully spin-polarized current through
the device. The fact that reaching $|\eta|=1$ in this manner---by varying
$\veps_2$ while dot 1 is held at particle-hole symmetry ($\veps_1=0$)---
is impossible to achieve in the noninteracting case $U_1=0$
indicates that the interference effects are more complex in the presence of
strong interactions.

It is important to emphasize that, in contrast to the maximal conductance value
$G_0$, the polarization $\eta$ is unaffected by asymmetry between the left and
right dot-lead couplings. Complete spin polarization ($|\eta|=1$) can be
achieved even in setups where $V_{2L}\ne V_{2R}$.

Figure \ref{fig:PolConducTeq0}(b) shows the variation of $\eta$ under field
sweeps at two different values $\veps_2>0$. For each position of the dot-2
level, $\eta$ changes sign at a nonzero $B$. For $\veps_2=0.02$, $\eta$ reaches
$+1$ at a small field and then dips to nearly $-1$ at a larger field before
increasing back toward zero. For $\veps_2=0.1$, by contrast, a small positive
peak in $\eta$ is followed at larger fields by a dip at (or very close to)
$-1$. This nonuniversal behavior reflects the subtlety of the interplay
between the field and particle-hole asymmetry in controlling the constructive
or destructive interference between transmission of electrons directly through
dot 2 and paths involving one or more detours to dot 1.

Similar ``spin-filtering'' effects in a magnetic field have been investigated
previously\cite{Aligia2004,Torio2004} in the context of a single-mode wire,
coupled near its midpoint via a tunnel junction to a quantum dot (the ``side
dot''). A number of experiments and models using different geometries for
spin-dependent transport have also been reported in the
literature.\cite{vdWiel,SpinPolRMP} Reference \onlinecite{Torio2004}
showed that conductance polarizations $\eta=1$ and $\eta=-1$ (in the language of
the present paper) occur at values of the dot energy $\veps_d(\eta=1)$ and
$\veps_d(\eta=-1)$ differing by a large scale exceeding the dot Coulomb
interaction strength $U$. Thus, the change in gate voltage needed to switch the
polarizations is so large that all traces of the Kondo effect are suppressed.
These behaviors should be contrasted with those found here, where the $\veps_2$
values that lead to $\eta=\pm 1$ differ only by an energy of order $\Delta_2$
(much smaller than $U_1$). What is more, the complete spin filtering achieved
in our setup depends crucially on the presence of Kondo many-body correlations.
This point will become particularly clear in the next section, where we
consider the effect of nonzero temperatures. Reference \onlinecite{Aligia2004}
considered a side-coupled quantum dot in regime of much smaller Kondo
temperatures. In contrast to our results for double quantum dots, complete
polarization of the conductance was reported to occur quite generically due to
a mechanism very similar to that we find in the noninteracting limit $U_1=0$
described by Eq.\ \eqref{eq:G:T=0:nonint}.

\subsection{Nonzero temperatures}

To this point, only zero-temperature results have been presented. This
subsection addresses the effect of finite temperatures on the zero-bias
conductance $G$ and its spin polarization $\eta$. Throughout the discussion,
temperatures are expressed as multiples of a characteristic many-body scale
$T_{K0}=0.021$, the system's Kondo temperature for $\veps_2=0$, $B=0$, and the
representative value $\lambda=0.0627$ that we have used in all our $T>0$
calculations.

\begin{figure}[t]
\includegraphics[width=3.3in]{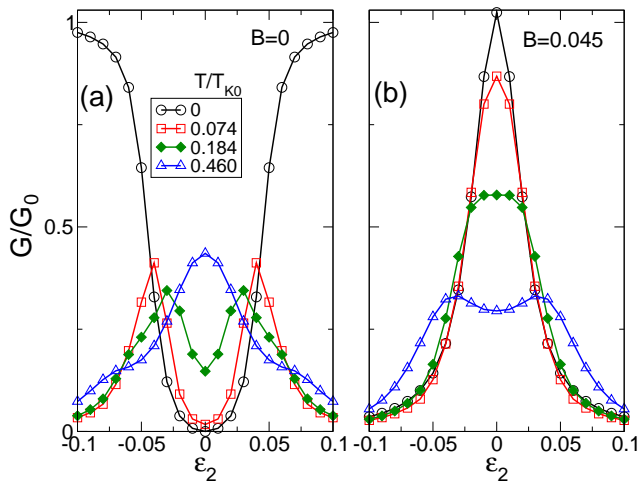}
\caption{\label{fig:FiniteT_Cond} (Color online)
Linear conductance $G$ vs dot-2 level position $\veps_2$ for
$U_1=-2\veps_1=0.5$ and $\lambda=0.0627$ at four temperatures $T$ for
(a) $B=0$, and (b) $B=0.045\simeq 2T_{K0}$.
Temperatures are expressed as multiples of $T_{K0}=0.021$.}
\end{figure}

Figure \ref{fig:FiniteT_Cond} plots $G$ vs $\veps_2$ for $\lambda=0.0627$
in fields $B=0$ (panel a) and $B\simeq B^{**} \simeq 2T_{K0}$ (panel b).
For $B=0$, the effect of increasing temperature is a progressive suppression of
the Kondo effect and hence of the conductance channel involving the many-body
Kondo resonance. As a result, $G$ rises near $\veps_2=0$ due to a lessening of
the destructive interference between the Kondo channel and the single-particle
resonance on dot 2 (discussed above in connection with Fig.\
\ref{fig:CondTeq0}), but there is a decrease in the conductance at
$|\veps_2|\gtrsim\Delta_2$, which is dominated by transmission through the Kondo
channel. This trend results in a conductance peak at some $|\veps_2|\ne 0$
for temperatures $0<T\lesssim T_{K0}$, which evolves into a peak centered at
$\veps_2=0$ for $T\gtrsim T_{K0}$, in which regime transmission is dominated by
the single-particle, Lorentzian-like contribution from dot 2.

Figure \ref{fig:FiniteT_Cond}(b) reveals a very different behavior for
$B=B^{**}\simeq 2T_{K0}$. As described above, the $T=0$ conductance attains its
maximum possible value $G_0$ at $\veps_2=0$ due to constructive interference
between the Kondo and single-particle conductance channels, and $G$ decreases
monotonically with increasing $|\veps_2|$. Raising the temperature over the
range $T\lesssim T_{K0}$ leads to suppression of the Kondo conductance channel
but has little effect on the single-particle channel, leading to a decrease in
$G$ that is strongest for $\veps_2=0$. Once the temperature passes $T_{K0}$, the
variation of $G$ with $\veps_2$ increasingly reflects the field splitting of
the dot-2 energy level, with peaks centered at $\veps_2\simeq \pm \half B$.

\begin{figure}[t]
\includegraphics[width=3.3in]{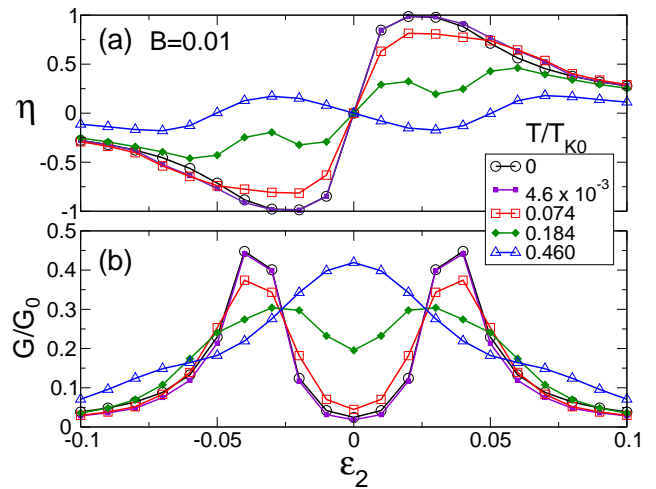}
\caption{\label{fig:FiniteT_Pol} (Color online)
(a) Conductance spin-polarization $\eta$, and (b) conductance $G$ vs dot-2
level energy $\veps_2$ at different temperatures for $U_1=-2\veps_1=0.5$,
$\lambda=0.0627$, and $B=0.01$. Temperatures are expressed as multiples of
$T_{K0}=0.021$.}
\end{figure}

The influence of temperature on the spin polarization of the conductance is
shown in Fig.\ \ref{fig:FiniteT_Pol}(a), which focuses on the case $B=0.01$
that we know from Fig.\ \ref{fig:PolConducTeq0} yields full spin polarization
($\eta=\pm 1$) at zero temperature for $\veps_2 \simeq \pm 0.025$. As $T$
increases from zero, the peak spin polarization is lowered, presumably due to a
combination of two effects: (i) a reduction in the destructive interference
between the Kondo and single-particle conduction channels for one spin species
$\sigma$ leading to an increase in
$-\mathrm{Im}\,\mathcal{T}_{\sigma}(\omega=0,T)$ entering Eq.\ \eqref{eq:G};
and (ii) thermal broadening of $-\partial f/\partial\omega$ in Eq.\
\eqref{eq:G} leading to sampling of $\omega$ values having nonzero
$-\mathrm{Im}\,\mathcal{T}_{\sigma}(\omega,T=0)$. At higher temperatures,
$T\simeq T_{K0}$, the suppression of the Kondo conductance channel unmasks
oscillations in $\eta$ vs $\veps_2$ that result from shifts in the
spin-resolved energy levels in dot 2. These oscillations are much less
pronounced than the polarization variations at lower temperatures and the
maximum values of $|\eta|$ are about an order of magnitude smaller than those
obtained in the Kondo regime.

Figure \ref{fig:FiniteT_Pol}(b) plots the total conductance $G$ vs $\veps_2$
corresponding to each of the $\eta$ vs $\veps_2$ traces in Fig.\
\ref{fig:FiniteT_Pol}(a). There is a close correlation (although not a perfect
match) between the $\veps_2$ values of the peaks in $G$ and of those in
$|\eta|$. This suggests that measurements of the total conductance can provide
a useful starting point for experiments seeking to optimize the system's
spin-filtering performance.

\begin{figure}[t]
\includegraphics[width=3.3in]{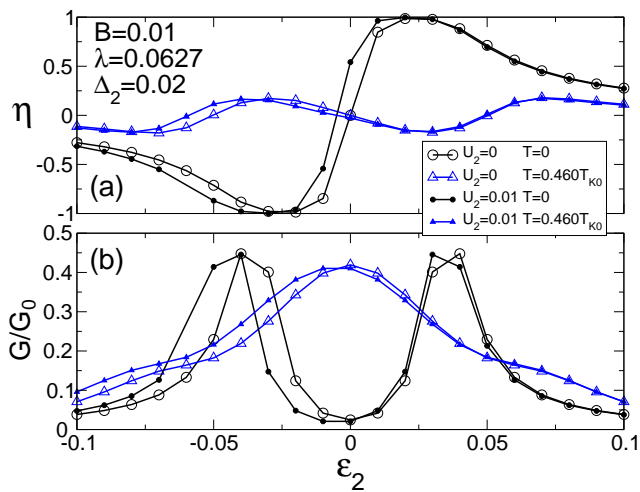}
\caption{\label{fig:U2neq0_FiniteT_Pol} (Color online)
Effect of a nonzero dot-2 interaction ($U_2>0$) on (a) the conductance
spin-polarization $\eta$, and (b) the conductance $G$, both plotted vs dot-2
level energy $\veps_2$ for the same parameters as in Fig.\
\ref{fig:FiniteT_Pol}. Open symbols correspond to $U_2=0$ and filled symbols to
$U_2=0.01$. Temperatures are expressed as multiples of $T_{K0}=0.021$.}
\end{figure}

Although we have focused on the special case $U_2=0$, the conductance features
described above by no means depend on this condition. In fact, qualitatively
similar results are obtained for an interacting dot 2 provided
that $U_2$ is small compared to the level broadening $\Delta_2$. This is
illustrated in Fig.\ \ref{fig:U2neq0_FiniteT_Pol}, which compares the $\veps_2$
dependence of the conductance and of its spin-polarization for $U_2=0$ (data
from Fig.\ \ref{fig:FiniteT_Pol}) and $U_2=\half\Delta_2=0.01$, both for the
lowest ($T=0$) and highest ($T=0.460 T_{K0}$) temperatures shown in Fig.\
\ref{fig:FiniteT_Pol}. Apart from a small shift in the point of particle-hole
symmetry, which moves from $\veps_2=0$ to $\veps_2=-\half U_2$, the other
essential features (such as the complete spin polarization at zero temperature)
are unaffected by the presence of Coulomb repulsion within dot 2.

\section{Conclusions}
\label{sec:conclusion}

In this work, we have investigated the effect of an applied magnetic field on
a strongly interacting quantum dot side-coupled to external leads via a weakly
interacting dot. Our numerical renormalization-group results show that the
interplay of electronic interference, the Kondo effect, and Zeeman splitting
brings about qualitative changes in the spectral and transport properties of
this system.
We have found, for instance, that the value of the interacting dot's
zero-temperature spectral function at the Fermi energy does not decay
monotonically with increasing field, as it does in single-dot setups.
Instead, the presence of the extra energy scale determined by the interdot
coupling introduces nonuniversal behavior, and in some cases leads to the
appearance of one or two maxima in the Fermi-energy spectral function at nonzero
values of $B$. These features can be understood by the presence of a
parameter-dependent phase appearing in the Friedel sum rule for
energy- and spin-dependent hybridization functions.

One of the signatures of the interplay of site and spin degrees of freedom in
this double-dot device is the appearance of spin-polarized currents between the
two leads. We have shown that the degree of spin polarization can be tuned up to
100\% by changing gate voltages and/or small magnetic fields in the system.
These results underscore the flexibility of quantum-dot systems for exploration
of novel effects in correlated electron physics.

\acknowledgments

We thank A.\ Seridonio and D.\ Logan for helpful discussions.
This work was supported in part under NSF Materials World Network Grants
DMR-0710540 and DMR-1107814 (Florida), and DMR-0710581 and DMR-1108285 (Ohio).
N.S., S.E.U., and E.V.\ acknowledge the hospitality of the KITP, and support
under NSF Grant PHY-0551164. L.D.S.\ acknowledges support from Brazilian
agencies CNPq (Grant No.\ 482723/2010-6) and FAPESP (Grant No.\ 2010/20804-9).
E.V.\ acknowledges support from CNPq (Grant No.\ 493299/2010-3) and FAPEMIG
(Grant No.\ CEX-APQ-02371-10). N.S. and S.E.U. acknowledge the hospitality of the
Dahlem Center and the support of the A. von Humboldt Foundation.

\end{document}